\documentclass[traditabstract,times]{aa} 

\newcommand{\ergps}{erg\thinspace s$^{-1}$}
\newcommand{\ergpspsqcm}{erg\thinspace s$^{-1}$\thinspace cm$^{-2}$}
\newcommand{\psqcm}{cm$^{-2}$}
\newcommand{\nH}{$N_{\rm H}$}

\usepackage{graphicx}
\begin{document}
\title{The XMM deep survey in the CDF-S II. a 9-20 keV selection of heavily obscured active galaxies at z$>$1.7}


   \author{K. Iwasawa\inst{1}\thanks{Email: kazushi.iwasawa@icc.ub.edu}
          \and
R. Gilli\inst{2}
\and
C. Vignali\inst{2,3}
\and
A. Comastri\inst{2}
\and
W.~N. Brandt\inst{4,5}
\and
P. Ranalli\inst{2,3,6}
\and
F. Vito\inst{2,3}
\and
N.~Cappelluti\inst{2}
\and
F.~J. Carrera\inst{7}
\and
S. Falocco\inst{7}
\and
I. Georgantopoulos\inst{6}
\and
V. Mainieri\inst{8}
\and
M. Paolillo\inst{9}
}

\institute{ICREA and Institut de Ci\`encies del Cosmos (ICC), Universitat de Barcelona (IEEC-UB), Mart\'i i Franqu\`es, 1, 08028 Barcelona, Spain
         \and
INAF - Osservatorio Astronomico di Bologna, Via Ranzani, 1, 40127 Bologna, Italy
\and
Universit\`a di Bologna - Dipartimento di Astronomia, Via Ranzani, 1, 40127 Bologna, Italy
\and
Department of Astronomy \& Astrophysics, 525 Davey Lab, The Pennsylvania State University, University Park, PA 16802, USA
\and
Institute for Gravitation and the Cosmos, The Pennsylvania State University, University Park, PA 16802, USA
\and
Institute of Astronomy \& Astrophysics, National Observatory of Athens, Palaia Penteli, 15236 Athens, Greece
\and
Instituto de F\'isica de Cantabria (CSIC-UC), 39005 Santander, Spain
\and
European Southern Observatory, Karl-Schwarzschild-Stra\ss e 2, 85748 Garching, Germany
\and
Dipartimento di Scienze Fisiche, Universit\`a di Napoli Fedelico II, C.U. di Monte Sant'Angelo, Via Cintia ed. 6, 80126 Napoli, Italy
          }


 
          \abstract{We present results on a search of heavily obscured
            active galaxies $z>1.7$ using the rest-frame 9-20 keV
            excess for X-ray sources detected in the deep XMM-CDFS
            survey. Out of 176 sources selected with the conservative
            detection criteria ($>8\sigma $) in the first source
            catalogue of Ranalli et al., 46 objects lie in the
            redshift range of interest with the median redshift
            $\tilde z\simeq 2.5$. Their typical rest-frame 10-20 keV
            luminosity is $10^{44}$ \ergps, as observed. Among
            optically faint objects that lack spectroscopic redshift,
            four were found to be strongly absorbed X-ray sources, and
            the enhanced Fe K emission or absorption features in their
            X-ray spectra were used to obtain X-ray spectroscopic
            redshifts. Using the X-ray colour-colour diagram based on
            the rest-frame 3-5 keV, 5-9 keV, and 9-20 keV bands, seven
            objects were selected for their 9-20 keV excess and were
            found to be strongly absorbed X-ray sources with column
            density of \nH $\thinspace\geq 0.6\times 10^{24}$ \psqcm,
            including two possible Compton thick sources. While they
            are emitting at quasar luminosity, $\sim 3/4$ of the
            sample objects are found to be absorbed by \nH $\thinspace
            > 10^{22}$ \psqcm. A comparison with local AGN at the
            matched luminosity suggests an increasing trend of the
            absorbed source fraction for high-luminosity AGN towards
            high redshifts. }

\keywords{X-rays: galaxies - Galaxies: active - Surveys
                             }
\titlerunning{9-20 keV excess sources in XMM-CDFS}
\authorrunning{K. Iwasawa et al.}
   \maketitle
%

\section{Introduction}

A population of heavily obscured Active Galactic Nuclei (AGN) at
cosmological distances, which might be missed by conventional quasar
surveys, has been postulated by AGN synthesis models of the X-ray
background (XRB, e.g., Gilli, Comastri \& Hasinger 2007; Treister,
Urry \& Virani 2009) and the super-massive black hole mass function in
the local Universe (Marconi et al 2004). Various infrared selections
have been employed extensively for seaching for these objects in which
strong re-radiation from obscuring dust is expected
(Mart\'inez-Sansigre et al 2005; Alonso-Herrero et al 2006; Daddi et
al 2007, Fiore et al 2008, 2009; Bauer et al 2010; Vignali et al 2010;
Alexander et al 2011; Luo et al 2011; Donley et al 2012). Although
X-ray observations should, in principle, also be effective for the
search on account of the intrinsic X-ray loudness of AGN (relative to
galaxy emission) and the penetrating power against obscuration, the
low throughput of the existing X-ray telescopes limits the
accessibility to high redshift. However, dedicated deep surveys with
extremely long exposures, for example, in the Chandra Deep Field South
(CDFS) conducted by XMM-Newton (Comastri et al 2011) and Chandra
(Giacconi et al 2002; Xue et al 2011) X-ray observatories now allow us
to pursue this approach. Here, we present a study of X-ray selected
heavily obscured active galaxies using the 3 Ms XMM-Newton survey of
CDFS.

X-ray absorption is measured by the low energy cut-off of an X-ray
spectrum, which moves to higher energies as absorbing
column density increases. When \nH\ approaches $10^{24}$ \psqcm, the cut-off
occurs above 10 keV. As demonstrated for nearby examples, such as NGC
4945 (Iwasawa et al 1993), the Circinus Galaxy (Matt et al 1999), and
NGC 6240 (Vignati et al 1999), detection of emission above 10 keV
plays a key role in discoveries of heavily obscured AGN in those
galaxies with absorbing column density exceeding $10^{24}$
\psqcm. This method works as long as the optical depth is not too
large, that is, when a source becomes fully Compton thick with \nH
$\thinspace\geq 10^{25}$ \psqcm, Compton down-scattering suppresses
the hard X-rays, leaving only reflected light, as observed in NGC 1068
(e.g., Matt et al 1997). While a direct access is not possible for
nearby objects with XMM-Newton, this crucial energy-band is redshifted
into its bandpass for high redshift objects at $z\geq 2$. Given the
shape of an absorbed X-ray spectrum, a negative K-correction sustains
the detectability of absorbed sources to high redshift. Utilizing
these properties, we searched for the rest-frame 9-20 keV excess
sources to identify heavily obscured AGN candidates in the sources
detected in the XMM-CDFS field.

The cosmology adopted here is $H_0=70$ km s$^{-1}$ Mpc$^{-1}$,
$\Omega_{\Lambda}=0.72$, $\Omega_{\rm M}=0.28$.

\section{The Sample}


\begin{figure}
\centerline{\includegraphics[width=0.4\textwidth,angle=0]{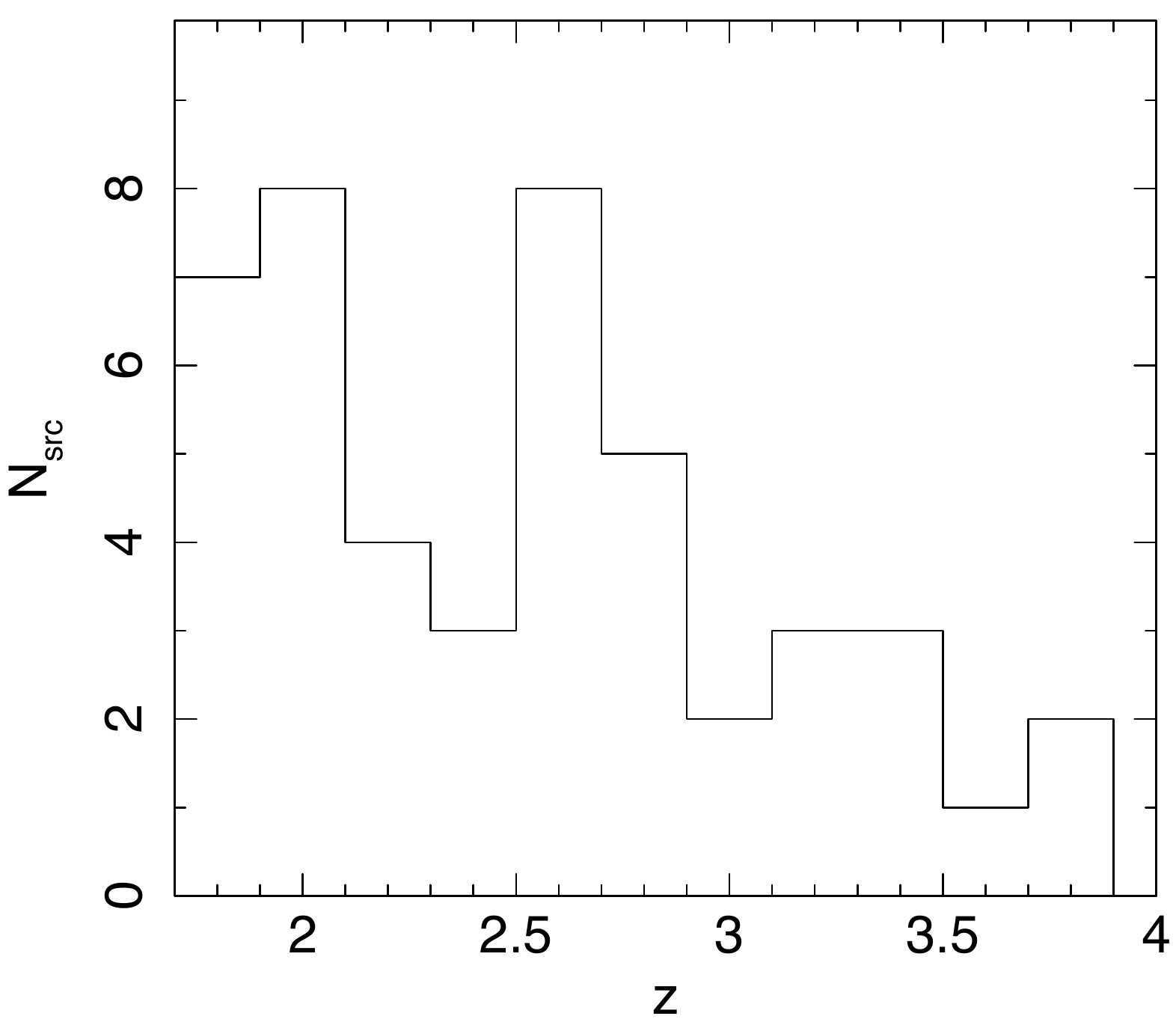}}
\caption{Distribution of redshifts of the 46 sources in the
  sample. PID 352 with X-ray determined redshift $z_{\rm X}=1.60$ is
  excluded. }
\end{figure}


\begin{figure}
\centerline{\includegraphics[width=0.4\textwidth,angle=0]{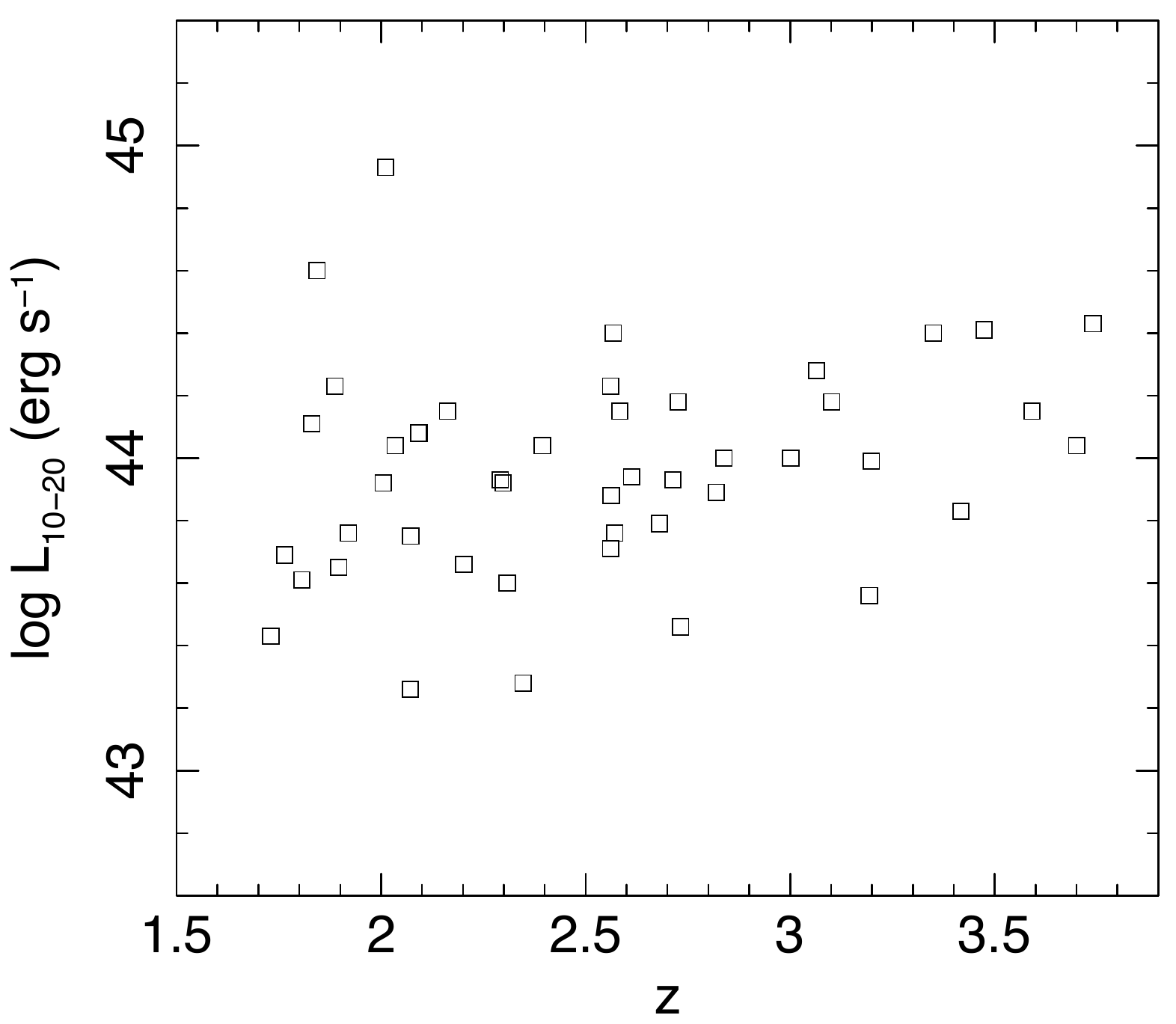}}
\caption{The distribution of the 46 objects in our sample in the
  rest-frame 10-20 keV luminosity vs. redshift plane. No
  correction for internal absorption has been made when calculating
  the luminosities. The median $L_{\rm 10-20}$ is $0.9\times 10^{44}$
  \ergps. }
\end{figure}

We selected sources from the first XMM-CDFS catalogue with
conservative detection criteria: 176 sources that were detected at
significance larger than $8\sigma $ in the 2-10 keV band and have
X-ray spectra verified for use for a spectral analysis are available
(details will be described in Ranalli et al in prep.). Since the
signal to noise ratio of individual spectra, obtained from the EPIC
cameras of XMM-Newton, falls steeply above 7.5 keV, we set the lower
bound of the redshift range of our sample to $z=1.7$, for which
rest-frame 20 keV corresponds to observed-frame 7.4 keV.

There are 47 objects with $z>1.7$ for which spectral data are
available from all the three EPIC cameras, pn, MOS1 and MOS2, apart
from two objects which are located outside the field of view of the pn
but within the two MOS cameras (see Table 1). Spectroscopic redshifts
are available for 33 objects, while photometric redshifts were
estimated by various papers (Luo et al 2008; Cardamone et al 2008;
Rafferty et al 2011; Wuyts et al 2008; Santini et al 2009; Taylor et
al 2009) for other 14 objects\footnote{The photometric redshift
  adopted in this paper are taken from Ranalli et al., in prep., in
  which the choices among various photometric redshift estimates are
  described in detail. }.  

For some objects with photometric redshifts, more constrained
redshift estimates could be obtained using the Fe K feature in their
X-ray spectra when they have a strongly absorbed X-ray spectrum. We
use these X-ray redshifts for five sources (Sect. 3.2). As a result
of X-ray redshift determination, one source ($z=1.60$), which had the
original photometric redshift $z=1.78$, went out of the redshift
range. This source (PID 352) is therefore excluded from the sample and
will not be discussed further.

Hereafter we use ``PID'' for the identification number of X-ray
sources listed in Ranalli et al, and the basic information of the 46
objects in the sample is presented in Table 1. The redshift
distribution of the sample is shown in Fig. 1. The median redshift is
$\tilde z=2.5$. The background-corrected counts obtained from the the
sum of the three EPIC cameras range from 400 to 8000 in the respective
rest-frame 3-20 keV band, while the typical counts are $\sim
1400$. The typical source fraction of the total (source plus
background) counts is $\simeq 0.4$ in both EPIC pn and MOS cameras.

Since the exposure time for each source varies, the observed flux in
the observed-frame 1-4 keV band, which is shared by all the sources
with various redshifts, is given in Table 1 as an objective
measure of source brightness. Median values of the observed frame 1-4
keV flux, $f_{1-4}$, the rest-frame 2-10 keV and 10-20 keV
luminosities, $L_{2-10}$, and $L_{10-20}$, are $2.5\times 10^{-15}$
\ergpspsqcm, $9.1\times 10^{43}$ \ergps, and $8.7\times 10^{43}$
\ergps, respectively. These luminosities are corrected for the
Galactic extinction, \nH $\thinspace = 9\times 10^{19}$ \psqcm (Dickey
\& Lockman 1990). Fig. 2 shows how the objects in our sample are
distributed in the $L_{10-20}$ - $z$ plane. The spread of the 10-20
keV luminosity is relatively narrow with a logarithmic dispersion of
0.3 (or a factor of $\sim 2$).


\begin{table*}
\begin{center}
\caption{Properties of the sample.}
\begin{tabular}{rcccccrccccccc}
PID & RA & Dec. & $z$ & & & Net & $s/m$ & $h/m$ & & $f_{1-4}$ & $L_{2-10}$ & $L_{10-20}$ \\
(1) & (2) & (3) & (4) & (5) & (6) & (7) & (8) & (9) & (10) & (11) & (12) & (13) \\[5pt]
26 & 53.21484 & $-27.97884$ & 3.198 & sp & a & 618 & $0.35 \pm 0.17$ &$0.97 \pm 0.33$ & A & 1.7e-15 & 9.2e+43 & 9.7e+43\\
30 & 53.03737 & $-27.97493$ & 1.830 & x & - & 1344 & $0.29 \pm 0.08$ &$2.06 \pm 0.27$ & V & 2.5e-15 & 4.5e+43 & 1.3e+44\\
31 & 53.28854 & $-27.97376$ & 2.583 & sp & a & 789 & $1.03 \pm 0.15$ &$1.07 \pm 0.19$ & M & 3.9e-15 & 2.1e+44 & 1.4e+44\\
33 & 53.25708 & $-27.97188$ & 1.843 & sp & a,b & 8112 & $1.20 \pm 0.04$ &$0.77 \pm 0.05$ & U & 2.5e-14 & 6.8e+44 & 4.0e+44\\
49 & 52.97654 & $-27.94723$ & 2.298 & sp & c & 1474 & $1.09 \pm 0.14$ &$1.07 \pm 0.18$ & M & 2.8e-15 & 1.0e+44 & 8.4e+43\\
57 & 53.29041 & $-27.93768$ & 2.571 & sp & a & 1891 & $1.49 \pm 0.14$ &$0.68 \pm 0.12$ & U & 2.8e-15 & 1.8e+44 & 5.7e+43\\
62 & 53.30284 & $-27.93094$ & 2.561 & sp & a & 3642 & $1.40 \pm 0.08$ &$0.60 \pm 0.07$ & U & 6.1e-15 & 3.8e+44 & 1.7e+44\\
64 & 53.17001 & $-27.92967$ & 3.350 & x & - & 1277 & $0.23 \pm 0.11$ &$1.95 \pm 0.31$ & V & 2.5e-15 & 8.7e+43 & 2.5e+44\\
68 & 53.25377 & $-27.92238$ & 2.005 & sp & d & 3023 & $1.30 \pm 0.10$ &$0.75 \pm 0.11$ & U & 5.5e-15 & 2.0e+44 & 8.4e+43\\
81 & 52.93824 & $-27.90995$ & 1.887 & sp & d & 5524 & $1.37 \pm 0.07$ &$1.03 \pm 0.09$ & U & 1.1e-14 & 3.1e+44 & 1.7e+44\\
84 & 53.06075 & $-27.90602$ & 2.561 & sp & e & 1226 & $0.99 \pm 0.19$ &$1.54 \pm 0.31$ & M & 1.3e-15 & 5.9e+43 & 5.1e+43\\
92 & 53.02137 & $-27.89840$ & 3.417 & ph & f & 1066 & $0.90 \pm 0.13$ &$0.91 \pm 0.16$ & M & 1.1e-15 & 9.0e+43 & 6.7e+43\\
93 & 53.00222 & $-27.89788$ & 2.819 & ph & g & 995 & $0.65 \pm 0.13$ &$1.05 \pm 0.22$ & M & 1.5e-15 & 7.2e+43 & 7.7e+43\\
97 & 53.19526 & $-27.89280$ & 2.732 & ph & h & 441 & $0.18 \pm 0.12$ &$0.75 \pm 0.24$ & A & 6.8e-16 & 3.0e+43 & 2.9e+43\\
98 & 53.01620 & $-27.89159$ & 3.001 & ph & f & 1411 & $0.88 \pm 0.13$ &$0.79 \pm 0.15$ & M & 1.3e-15 & 9.3e+43 & 1.0e+44\\
103 & 52.94689 & $-27.88717$ & 2.034 & ph & g & 3901 & $1.17 \pm 0.07$ &$0.94 \pm 0.08$ & U & 6.4e-15 & 2.2e+44 & 1.1e+44\\
107 & 52.20927 & $-27.88088$ & 3.474 & sp & c & 2468 & $0.78 \pm 0.06$ &$1.01 \pm 0.07$ & M & 3.7e-15 & 2.9e+44 & 2.6e+44\\
108 & 53.05395 & $-27.87688$ & 2.562 & sp & e & 1164 & $0.45 \pm 0.06$ &$1.19 \pm 0.13$ & A & 1.6e-15 & 5.7e+43 & 7.5e+43\\
114 & 53.32027 & $-27.87144$ & 1.806 & sp & a & 508 & $0.09 \pm 0.20$ &$1.92 \pm 0.61$ & V & 1.0e-15 & 2.2e+43 & 4.1e+43\\
116 & 53.04742 & $-27.87028$ & 3.740 & x & - & 2176 & $0.45 \pm 0.05$ &$1.18 \pm 0.08$ & A & 3.0e-15 & 1.9e+44 & 2.7e+44\\
120 & 53.17388 & $-27.86740$ & 3.591 & sp & e & 1987 & $1.24 \pm 0.10$ &$0.95 \pm 0.09$ & U & 2.1e-15 & 2.5e+44 & 1.4e+44\\
144 & 53.12423 & $-27.85159$ & 3.700 & sp & e & 752 & $0.29 \pm 0.16$ &$1.46 \pm 0.33$ & V & 1.0e-15 & 4.2e+43 & 1.1e+44\\
158 & 52.96872 & $-27.83828$ & 2.394 & sp & e & 1331 & $0.65 \pm 0.09$ &$1.34 \pm 0.15$ & M & 2.4e-15 & 8.2e+43 & 1.1e+44\\
173 & 53.18078 & $-27.82065$ & 1.920 & sp & e & 2951 & $1.14 \pm 0.07$ &$0.80 \pm 0.07$ & U & 3.5e-15 & 1.1e+44 & 5.7e+43\\
180 & 53.16528 & $-27.81382$ & 3.064 & sp & e & 1427 & $0.22 \pm 0.05$ &$2.01 \pm 0.18$ & V & 2.2e-15 & 7.4e+43 & 1.9e+44\\
190 & 53.26092 & $-27.80650$ & 3.101 & ph & f & 1965 & $1.05 \pm 0.08$ &$0.87 \pm 0.08$ & M & 2.8e-15 & 2.1e+44 & 1.5e+44\\
194 & 53.03947 & $-27.80214$ & 2.838 & sp & b,e & 1464 & $0.49 \pm 0.05$ &$1.03 \pm 0.08$ & A & 2.2e-15 & 1.1e+44 & 1.0e+44\\
200 & 53.24949 & $-27.79664$ & 2.567 & sp & e & 6545 & $1.35 \pm 0.05$ &$0.75 \pm 0.04$ & U & 8.1e-15 & 5.1e+44 & 2.5e+44\\
201 & 52.91637 & $-27.79593$ & 2.713 & sp & d & 1457 & $1.42 \pm 0.13$ &$0.78 \pm 0.11$ & U & 2.6e-15 & 2.0e+44 & 8.6e+43\\
210 & 53.17847 & $-27.78400$ & 3.193 & sp & e & 1106 & $1.04 \pm 0.16$ &$0.69 \pm 0.17$ & M & 9.5e-16 & 8.7e+43 & 3.6e+43\\
211 & 53.03384 & $-27.78214$ & 2.612 & sp & c & 1675 & $1.07 \pm 0.10$ &$1.01 \pm 0.12$ & M & 2.3e-15 & 1.1e+44 & 8.7e+43\\
213 & 53.27462 & $-27.78054$ & 2.202 & ph & f & 1008 & $0.92 \pm 0.17$ &$0.99 \pm 0.25$ & M & 1.4e-15 & 4.6e+43 & 4.6e+43\\
215 & 53.02193 & $-27.77877$ & 2.071 & sp & c & 690 & $0.12 \pm 0.08$ &$0.86 \pm 0.16$ & A & 1.1e-15 & 2.7e+43 & 4.0e+43\\
219 & 52.95665 & $-27.77598$ & 2.308 & sp & d & 1042 & $1.65 \pm 0.30$ &$0.96 \pm 0.27$ & U & 1.6e-15 & 7.5e+43 & 4.0e+43\\
231 & 53.09398 & $-27.76715$ & 1.730 & sp & d & 1111 & $0.64 \pm 0.09$ &$1.15 \pm 0.18$ & M & 1.2e-15 & 2.4e+43 & 2.7e+43\\
245 & 53.08279 & $-27.75493$ & 2.680 & x & - & 373 & $0.08 \pm 0.13$ &$2.23 \pm 0.54$ & V & 7.9e-16 & 2.3e+43 & 6.1e+43\\
252 & 53.08340 & $-27.74644$ & 1.896 & sp & d & 441 & $0.09 \pm 0.15$ &$2.44 \pm 0.59$ & V & 7.8e-16 & 1.9e+43 & 4.5e+43\\
269 & 53.37135 & $-27.73200$ & 1.764 & ph & g & 663 & $0.93 \pm 0.16$ &$1.20 \pm 0.28$ & M & 2.8e-15 & 5.9e+43 & 4.9e+43\\
283 & 53.10732 & $-27.71837$ & 2.291 & sp & e & 2076 & $0.84 \pm 0.08$ &$1.10 \pm 0.12$ & M & 2.7e-15 & 8.7e+43 & 8.6e+43\\
285 & 53.28673 & $-27.71504$ & 2.072 & sp & d & 1798 & $1.37 \pm 0.14$ &$0.76 \pm 0.15$ & U & 3.3e-15 & 1.2e+44 & 5.6e+43\\
302 & 53.39597 & $-27.70256$ & 2.011 & sp & a & 4054$^{\dagger}$  &$1.19 \pm 0.05$ &$0.83 \pm 0.06$ & U & 4.3e-14 & 1.4e+45 & 8.6e+44\\
311 & 53.25599 & $-27.69488$ & 2.091 & sp & i & 3351 & $1.08 \pm 0.07$ &$1.06 \pm 0.10$ & M & 5.7e-15 & 1.8e+44 & 1.2e+44\\
327 & 53.37052 & $-27.67841$ & 2.162 & sp & d & 638 & $0.74 \pm 0.12$ &$0.63 \pm 0.16$ & M & 5.0e-15 & 1.8e+44 & 1.4e+44\\
332 & 53.37428 & $-27.66908$ & 2.092 & sp & d & 510$^{\dagger}$  &$0.78 \pm 0.12$ &$0.51 \pm 0.15$ & M & 5.6e-15 & 1.7e+44 & 1.2e+44\\
366 & 53.07546 & $-27.61604$ & 2.347 & ph & g & 882 & $1.47 \pm 0.26$ &$1.37 \pm 0.30$ & U & 1.6e-15 & 7.6e+43 & 1.9e+43\\
503 & 53.00248 & $-27.72286$ & 2.726 & sp & e & 2330 & $0.96 \pm 0.06$ &$0.82 \pm 0.06$ & M & 3.8e-15 & 2.3e+44 & 1.5e+44\\[5pt]
\end{tabular}
\begin{list}{}{}
  Note --- (1) Source identification number in the XMM-CDFS catalogue
  of Ranalli et al; (2),(3) XMM position of source (degrees, J2000);
  (4) redshift; (5) source of redshift, sp: optical spectroscopic; ph:
  photometric; x: X-ray spectroscopic; (6) references for the redshift
  estimates: a: Treister et al (2009b); b: Cooper et al (2011); c:
  Popesso et al (2009); d: Silverman et al 2010; e: Szokoly et al
  (2004); f: Luo et al (2008); g: Cardamone et al (2008); h: Rafferty
  et al (2011); and i: Balestra et al (2010). (7) net counts in the
  rest-frame 3-20 keV band from all the three EPIC
  cameras. $\dagger$: MOS1+MOS2 only; (8) X-ray colour, $s/m$: photon
  ratio of the rest-frame 3-5 keV and 5-9 keV; (9) X-ray colour,
  $h/m$, photon ratio of the rest-frame 9-20 keV and 5-9 keV; (10)
  X-ray colour category; (11) observed-frame 1-4 keV flux in units of
  \ergpspsqcm ; (12) rest-frame 2-10 keV luminosity in units of \ergps
  ; (13) rest-frame 10-20 keV luminosity in units of \ergps. The
  luminosities given here are corrected for the Galactic absorption.
\end{list}
\end{center}
\end{table*}

\section{Results}

\subsection{X-ray colour analysis}


\begin{figure}
\centerline{\includegraphics[width=0.45\textwidth,angle=0]{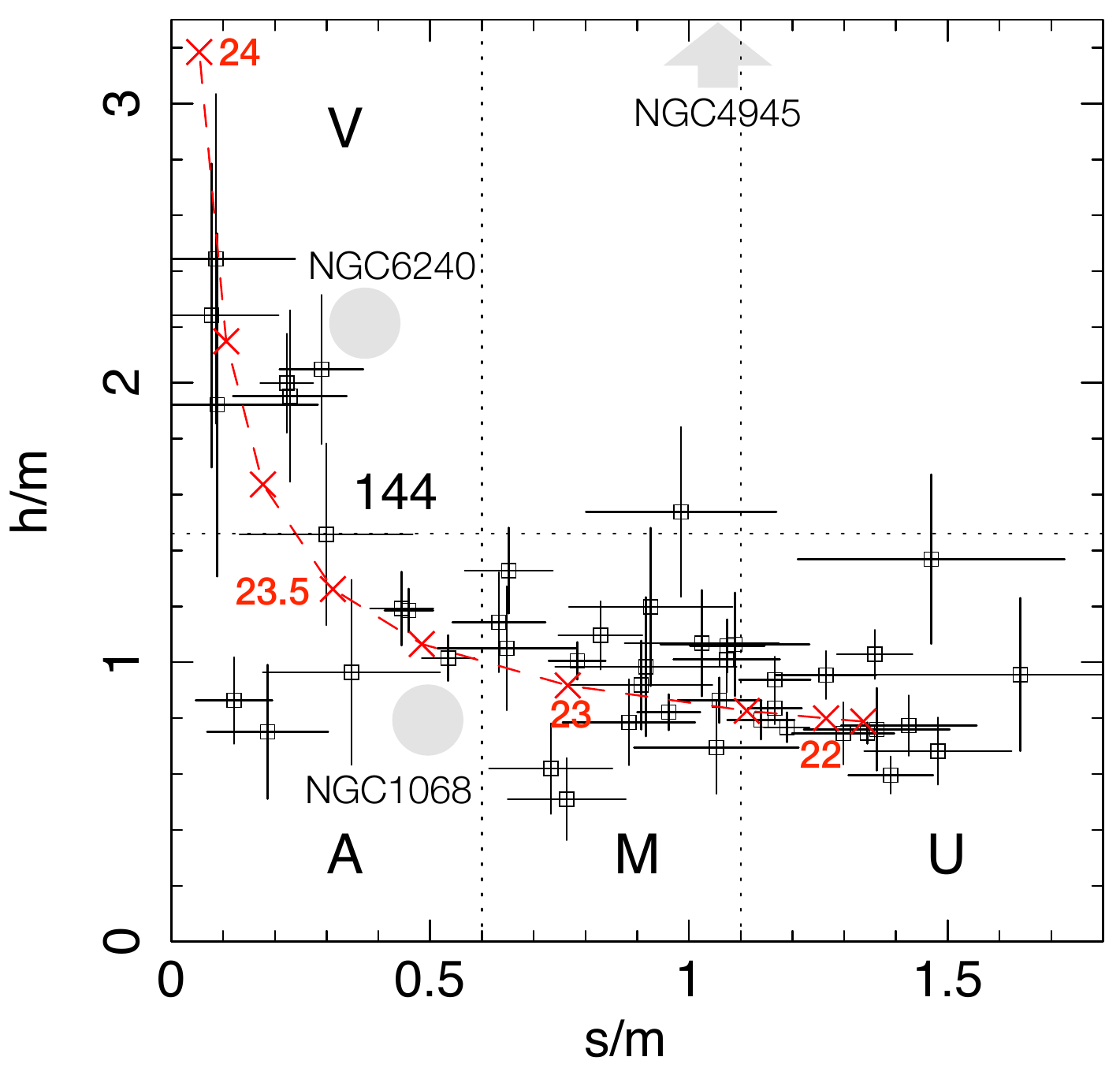}}
\caption{The X-ray colour-colur diagram, based on the data obtained
  from the XMM-Newton EPIC cameras, where $s$, $m$ and $h$ are
  the detector-response-corrected photon counts in the rest-frame
  bands of 3-5 keV, 5-9 keV and 9-20 keV, respectively. The four
  categories, V, A, M and U and their boundaries are indicated. Our
  reference heavily obscured AGN, PID 144, is labeled in the
  diagram. The red dashed-line indicates the evolution track of the
  X-ray colour when a power-law of $\Gamma = 1.8$ is modified by
  various absorbing colmun. The crosses mark log $N_{\rm H}$ values
  21, 22, 22.5, 23, 23.3, 23.5, 23.7, 23.85 and 24 (\psqcm) from the
  bottom-right to the upper-left along the track. The X-ray colours
  estimated for the nearby, heavily obscured AGN, NGC 6240, NGC 4945
  and NGC 1068 are also plotted. Note that NGC 4945 has a large value
  of $h/m=6.8$, which is outside of the frame (see text for details of
  these sources). X-ray spectra of the two sources (PID 84 and 366)
  with $h/m\sim 1.4$, located in the M and U intervals, respectively,
  are described in text (Sect. 3.1).}
\end{figure}


\begin{figure}
\centerline{\includegraphics[width=0.35\textwidth,angle=0]{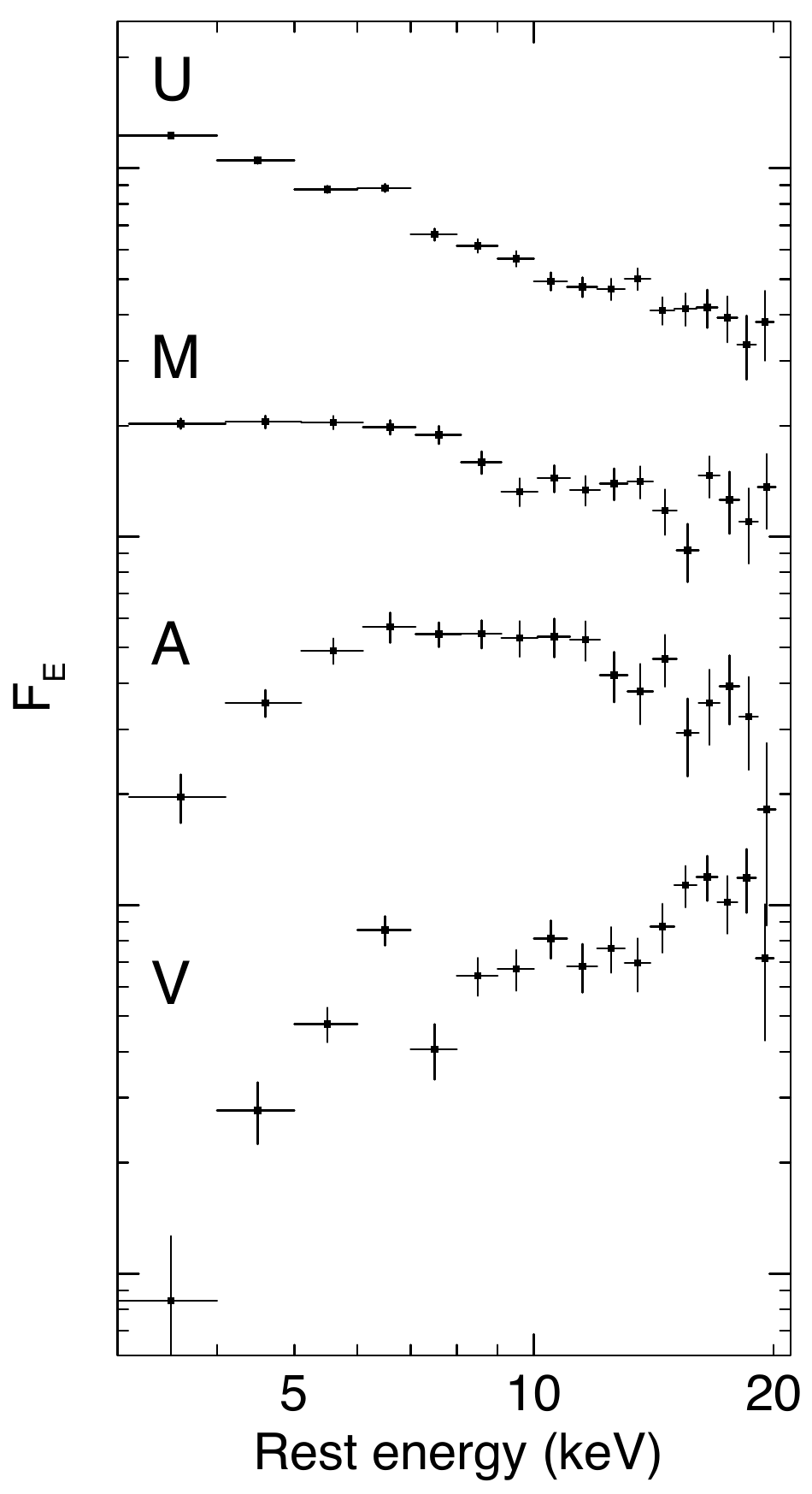}}
\caption{The rest-frame 3-20 keV stacked spectra for the four
  categories, defined in Fig. 3. The vertical axis is in arbitrary
  unit of flux density. Only the XMM-Newton data were used. The
  spectral stacking is a straight sum of individual sources while a
  weighted mean of the available EPIC data, based on the signal to
  noise ratio, is taken for each source. Number of sources, typical
  redshift and luminosity of each category can be found in Table
  2. For a reference, the spectral slope of the U category spectrum is
  $\alpha\simeq 1.8$, where $F_{\rm E}\propto E^{-\alpha}$, i.e.,
  photon index $\Gamma\simeq 1.8$. }
\end{figure}

\begin{table}
\begin{center}
\caption{Properties of the four X-ray colour categories, V, A, M, and U.}
\begin{tabular}{crcccc}
Category & N & $\tilde z$ & log $\tilde L_{2-10}$ & log $\tilde L_{10-20}$ & $\Gamma_{10-20}$ \\
(1) & (2) & (3) & (4) & (5) & (6)\\[5pt]
V & 7 & 2.68 & 43.59 & 44.04 & $0.3\pm 0.2$ \\
A & 6 & 2.78 & 43.86 & 43.93 &  $1.5\pm 0.3$ \\
M & 19 & 2.56 & 43.96 & 43.93 & $1.3\pm 0.2$ \\
U & 14 & 2.19 & 44.20 & 43.99 & $1.6\pm 0.1$  \\
\end{tabular}
\begin{list}{}{}
\item[] Note --- (1) The X-ray colour category (see Fig. 3); (2)
  Number of objects belonging to the category; (3) Median redshift;
  (4) Median rest-frame 2-10 keV luminosity; (5) Median rest-frame
  10-20 keV luminosity; (6) Photon index in the 10-20 keV band of the
  stacked spectrum (see Fig. 3). Photon index $\Gamma $ is related to
  the energy index $\alpha $, where flux density, $F_{\rm E}\propto
  E^{-\alpha}$, by $\Gamma = \alpha + 1$.
\end{list}
\end{center}
\end{table}

For selecting sources with various degrees of absorption, three rest-frame
energy bands: $s$ (3-5 keV); $m$ (5-9 keV); and $h$ (9-20 keV), are
defined and two X-ray colours: $s/m$ and $h/m$ are computed. At
energies above 3 keV, little contribution from soft X-ray emission
originating from the extranuclear region is expected. As the intrinsic
continuum slope in the 3-20 keV band is not expected to vary wildly
between objects, absorption would be the main driver of changes in the
X-ray colours. For the adopted rest-frame energy range, these X-ray
colours are sensitive to column densities larger than \nH
$\thinspace\simeq 10^{22}$ \psqcm.

Since our objects have a wide range of redshift (1.7-3.8 in $z$),
these X-ray colours are derived using photon spectra, i.e., spectral
data corrected for the detector response and the Galactic absorption
as a function of the rest-frame energy. The correction method employed
here is practically the same as that used in the XMM-COSMOS
spectral stacking analaysis (Iwasawa et al 2012). The photon counts in
each band are the weighted mean of the three EPIC cameras, where we
adopted the signal-to-noise ratio in the rest-frame 3-20 keV band as
the weight. The two X-ray colours, $s/m$ and $h/m$, for individual
sources are listed in Table 1, and the colour-colour diagram is shown
in Fig. 3.

With the two X-ray colours, a column density range of log
\nH\ = 22-24 (\psqcm ) can be probed, as $s/m$ covers the lower \nH\
regime and $h/m$ does the higher. In Fig. 3, a locus of spectral
evolution when a power-law continuum of photon index $\Gamma = 1.8$ is
modified by various absorbing column of log $N_{\rm H}$ between 21 and
24 (\psqcm ) is drawn. As the $s/m$ represents softness of a spectrum
below 9 keV, objects at the bottom-right in Fig. 3 are populated by
sources with little absorption. The $s/m$ colour moves to the left as
absorption increases. Two divisions were made along the $s/m$ axis, at
$s/m = 0.6$ and 1.1. In the lowest interval, the model locus turns
upwards as increasing absorption at log \nH $\geq 23.5$ (\psqcm ) and
a few sources indeed spread towards higher h/m values, which indicates an
excess of 9-20 keV emission. 

PID 144 ($z=3.70$) is a previously known, heavily obscured AGN with
an X-ray absorbing column of \nH $\thinspace\sim $(0.6-0.9)$\times
10^{24}$ \psqcm (Norman et al 2002; Comastri et al 2011), located in
this interval. We take this object with $h/m=1.46$ as the reference
and sources that have $h/m$ larger than this object were classified as
9-20 keV excess sources. 

According to the three intervals along $s/m$ and two intervals along
$h/m$, four zones, V: Very absorbed; A: Absorbed; M: Modestly
absorbed; and U: Unabsorbed, are defined in the colour-colour diagram,
as shown in Fig. 3. The degree of absorption thus increases in the
order of U, M, A, and V, and typical column densities for these X-ray
colour categories would be log $N_{\rm H}$ of $\leq 22$, 22.7, 23.4,
and 23.8 (\psqcm ), respectively.

For a comparison, the X-ray colours of nearby, well-studied heavily
obscured AGN, NGC 6240 ($N_{\rm H}\sim 2\times 10^{24}$ \psqcm), NGC
4945 ($N_{\rm H}\sim 5\times 10^{24}$ \psqcm), and NGC 1068 ($N_{\rm
  H}\geq 10^{25}$ \psqcm) were computed, based on the spectra
presented in Vignati et al (1999), Guainazzi et al (2002), and Matt et
al (1997), respectively, obtained from the BeppoSAX observations (see
Fig. 3). In these low luminosity systems, non AGN components, e.g., a
circumnuclear starburst, flaring X-ray binaries (e.g., Brandt, Iwasawa
\& Reynolds 1996), can make a significant contribution to their
spectra in the lower energy range, altering the $s/m$ colour in
particular, more than in high luminosity AGN like our sample. Despite
of this spectral complexity, $h/m$ serves as a good indicator of
strongly absorption seen in sources like NGC 6240 and NGC 4945. The
$h/m$ colour moves back to a lower value for a fully Compton thick
source, e.g., NGC 1068, but it still remains in a zone of hard
spectrum sources.

Two sources, PID 84 and PID 366, have $h/m$ values similar to the
reference PID 144 but softer $s/m$ colours (see Table 1). An
inspection of their spectra shows that PID 84 has a moderately
absorbed spectrum with \nH $\thinspace\simeq 1\times 10^{23}$ \psqcm\
as expected for the M category, while PID 366 in the U interval shows
a relatively soft spectrum but with a deficit at the rest-frame 7-10
keV (observed 2.2-3 keV range), causing the large value of $h/m$. This
could be attributed to a strong Fe K edge caused by absorption of \nH
$\thinspace\sim 6\times 10^{23}$ \psqcm, where a spectral complexity
might play a role to mask the strong absorption.

The sources in the V and A categories are absorbed by \nH\ of a few
times of $10^{23}$ \psqcm\ or larger, so that prominent Fe K features
in the form of an emission line or an absorption edge can be
observed. This offers a possibility to derive a reliable X-ray
spectroscopic redshift. There are five objects in the two categories
with only photometric redshifts. X-ray redshift ($z_{\rm X}$) were
obtained for these five objects and their X-ray colours were
recomputed assuming the new redshifts. Details of the X-ray redshift
measurements are described in Sect. 3.2. 

Basic information on the sources in the four categories is given in
Table 2 and their stacked, rest-frame 3-20 keV spectra are shown in
Fig. 4, which demonstrates representative spectral shapes for
respective categories.  
Note that exceptionally hard 10-20 keV spectrum of V compared to the
other three, indicating that large absorption column (\nH
$\thinspace\sim 10^{24}$ \psqcm ) are affecting the sources in this
category (Table 2).

\subsection{X-ray redshift measurements}

Redshift was measured with X-ray spectra for five objects, PID 30, 64,
116, 245, and 352, which have only photometric redshifts (Table
3). These objects are too faint in the optical band
to obtain a reliable spectroscopic redshift. The photometric
redshifts reported by various authors for each object spread over a
significant range, while they can serve as a guide for a redshift
range to be searched in. The Fe K features imprinted in their absorbed
spectra gave improved accuracy in the redshift measurements. The
Chandra data from the 4 Ms (Xue et al 2011) and the ECDFS (Lehmer et
al 2005) observations were also added to the analysis for improving
the spectral quality. The observed-frame 0.5-7 keV spectra of these
sources are shown in Fig. 5, except for PID 352, details of which will
be reported in a separate paper. They are photon spectra combining
XMM-Newton and Chandra data for displaying purpose only. All the
spectral results presented hereafter were obtained by fitting spectral
datasets from different cameras jointly.

The redshift determination was principally driven by the Fe K edge,
which is assumed to arise from cold medium and thus at the energy of
7.1 keV, as it is normally statistically more robust feature than the
line. Two exceptions are PID 116 and PID 352, for which the Fe K
emission-lines, detected at $2.5\sigma $ with EW $\approx 0.15$ keV
and $4.5\sigma $ with EW $\approx 0.30$ keV, respectively, were used
to obtain the redshifts assuming the rest-frame line energy of 6.4 keV
emitted from cold matter.

This assumption may not be true for PID 116, which is a luminous
submillimetre galaxy detected at 870 $\mu $m with LABOCA (LESS 9,
Wardlow et al 2011; Biggs et al 2011). As the X-ray detected luminous
infrared galaxies with $L_{\rm IR}\sim 10^{13} L_{\odot}$ at $z>2$ in
the COSMOS field appears to show high-ionization Fe K emission, e.g.,
Fe {\sc xxv} at 6.70 keV and/or Fe {\sc xxvi} at 6.97 keV as inferred
from a spectral stacking analysis (Iwasawa et al 2012), the emission
line of PID 116 could also be either of these high-ionization
lines. In this case, the redshift would be $z=3.96$ or $z=4.16$, when
it was identified with Fe {\sc xxv} and Fe {\sc xxvi},
respectively. The former value is close to the photometric redshift
derived by Luo et al (2008) and the latter to the secondary solution
of Wardlow et al (2011) (see Table 3).

As shown in Table 3, $z_{\rm X}$ are
found to lie within the range of various photometric redshifts.

\begin{table}
\begin{center}
\caption{X-ray redshift measurements.}
\begin{tabular}{rcl}
PID & $z_{\rm X}$ & Photo-z \\[5pt]
30 & $1.83\pm 0.07$ & $2.123^a, 1.936^b, 1.84^c, 1.683^d$  \\
64 & $3.35\pm 0.04$ & $3.528^a, 3.341^c, 3.301^e$ \\ 
116 & $3.74\pm 0.06$ & $3.53^a, 3.99^b, 4.63^f, 4.14^g$ \\
245 & $2.68\pm 0.12$ & $3.001^a, 2.431^h, 2.28^k$ \\ 
352 & $1.60\pm 0.02$ & $1.78^d$ \\
\end{tabular}
\begin{list}{}{}
\item[] Note --- $z_{\rm X}$ is the X-ray spectroscopic redshift with
  $1\sigma $ error. The Fe K features are assumed to arise from cold
  matter (see text for details).
\item[References for photometric redshifts] a: Luo et al (2008); b:
  Cardamone et al (2008); c: Rafferty et al (2011); d: Taylor et al
  (2009); e: Wuyts et al (2008); f: Wardlow et al (2011); g: Wardlow
  et al (2011, the second solution); h: Dahlen et al (2010); k:
  Santini et al (2009).
\end{list}
\end{center}
\end{table}

\begin{figure}
  \centerline{\includegraphics[width=0.47\textwidth,angle=0]{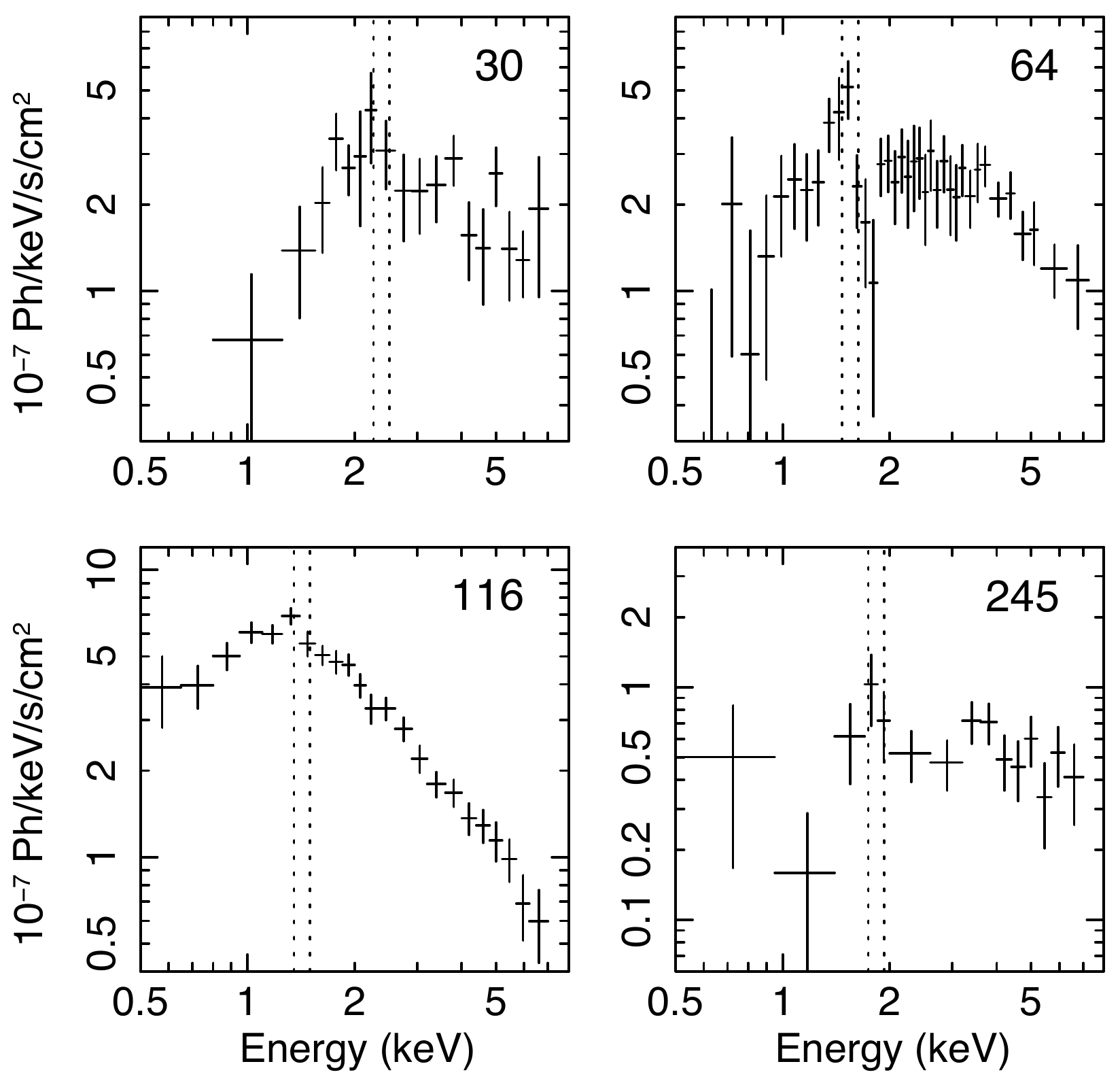}}
  \caption{The X-ray spectra of the four objects (PID 30, 64, 116,
    245) whose redshifts were determined using the Fe K features. The
    Chandra ACIS-I data, obtained in the deep CDFS (4 Ms) and the
    ECDFS observations, were combined with the XMM-Newton data. The
    rest-frame 6.4 keV and 7.1 keV which would be observed with those
    redshifts are indicated by the dotted lines. PID 30, 64, and 245
    are heavily obscured sources in the V category while PID 116 is a
    source in the A category.}
\end{figure}

\subsection{9-20 keV excess sources}

\begin{figure}
  \centerline{\includegraphics[width=0.47\textwidth,angle=0]{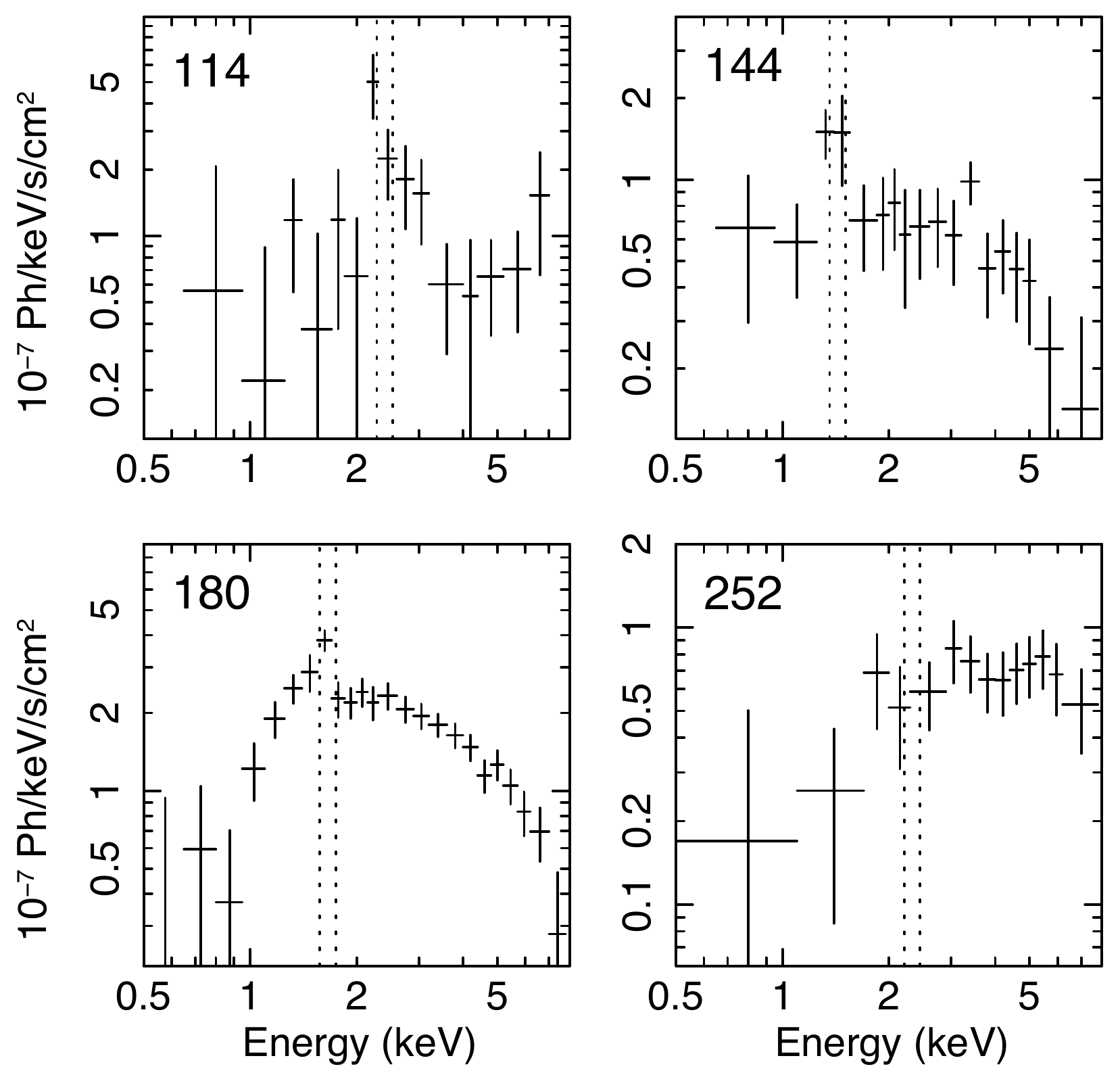}}
  \caption{The X-ray spectra of the four objects (PID 114, 144, 180,
    252) which complete the seven objects of the V category in
    addition to the three objects shown in Fig. 5. The rest-frame 6.4
    keV and 7.1 keV are indicated by the dotted lines that were
    computed assuming the spectroscopic redshifts for respective
    objects.}
\end{figure}

\begin{table}
\begin{center}
\caption{X-ray absorption in the 9-20 keV excess sources.}
\begin{tabular}{rccc}
PID & $N_{\rm H,24}$ & $AC_{2-10}$ & $AC_{10-20}$ \\
(1) & (2) & (3) & (4) \\[5pt]
30 & $0.57^{+0.07}_{-0.06}$ & 7 & 1.5 \\
64 & $0.83^{+0.05}_{-0.04}$ & 12 & 1.9 \\
114$^{\dagger}$ & $0.40^{+0.15}_{-0.10}$ & 4 & 1.3 \\
144 & $0.81^{+0.06}_{-0.06}$ & 11 & 1.9 \\
180 & $0.55^{+0.03}_{-0.02}$ & 6 & 1.5 \\
245 & $0.96^{+0.09}_{-0.08}$ & 15 & 2.1 \\ 
252$^{\dagger}$ & $0.97^{+0.11}_{-0.09}$ & 16 & 2.1 \\
\end{tabular}
\begin{list}{}{}
\item[] (1) Source identification number; (2) Absorption column
  density in unit of $10^{24}$ \psqcm; (3) Absorption correction
  factor for the 2-10 keV luminosity; (4) Absorption correction factor
  for the 10-20 keV luminosity.
\item[] Note --- Spectral fits were performed for the observed-frame
  1-7 keV data using an absorbed power-law with $\Gamma =1.8$. When a
  Thomson opacity approaches unity, as observed in these sources, the
  absorption correction depends on the geometry of absorbing clouds
  (e.g., Matt et al 1999). In this table, the values for a spherical
  geometry are given as the lower limits. These values can go up by a
  factor of $\sim 2$, as the covering factor of the absorber is
  reduced. $\dagger $: The spectra of these sources can also be
  described well by a reflection spectrum from cold matter.
\end{list}
\end{center}
\end{table}

The spectra of the seven objects in the V category are shown in Fig. 5
and 6. For displaying purpose, EPIC pn, EPIC MOS1, MOS2, and the
Chandra data from the ACIS detector are combined together. All the
spectra show spectral discontinuities at 6-7 keV, indicating an Fe K
line and/or a deep Fe K absorption edge, in agreement with strong
absorption. 

To estimate the absorbing column density of these sources, an absorbed
power-law model with photon index of $\Gamma = 1.8$ is fitted (Table
4). The absorption model with the Wisconsin cross section (Morrison \&
McCammon 1983) and the one with the effects of Compton scattering
taken into account, {\tt PLCABS} (Yaqoob 1997), give consistent
results on \nH. Whilst the absorption cut-off is slightly modified
when Compton scattering is taken into account, the 7-20 keV spectrum,
shaped by an Fe K edge, remains unchanged in shape for the $N_{\rm H}$
range of our sources, \nH $\thinspace\leq 10^{24}$\psqcm, (although
the flux is further suppressed). This also applies to recently
developed more sophisticated X-ray spectral models (e.g., Ikeda, Awaki
\& Terashima 2009; Murphy \& Yaqoob 2010). As our \nH\ fits are mainly
driven by the data in the Fe K edge band, it can be understood that
both absorption models gives similar \nH, given the data quality of
the spectra. However, since Compton scattering reduces the continuum
level further compared to the case where the scattering effect is not
taken into account, the absorption correction factor for estimating an
intrinsic continuum luminosity would be larger. This effect also
depends on the geometry of the absorber (e.g., Matt et al 1999) which
is not known for our objects. In Table 4, we give
absorption-correction factors for a spherical absorber. These factors
could go up by a factor of $\sim 2$ as the covering factors decreases
down to that of a disk-like geometry for the relevant range of \nH.

The column densities given in Table 4 were obtained, assuming the
observed emission is transmitted light through an absorber. However,
the hard X-ray colour exhibited by these objects could also result
from a reflection-dominated spectrum of a Compton thick source. This
is probably the case for PID 114, in which a strong Fe K line is
detected (see below and Table 5), whereas the apparently moderate
column density is inferred from the absorption model for the poor
quality continuum spectrum (Table 4). PID 252 has the hardest spectrum
in terms of the hard X-ray colour $h/m$ (Fig. 6) although no obvious
Fe line is seen. For these two objects, a pure reflection spectrum
from cold matter, modelled by {\tt pexrav} (Magdziarz \& Zdziarski
1995) or {\tt pexmon} (Nandra et al 2007), provides a comparable fit
to their spectra, compared to the absorption model. This indicates
that these two objects might be Compton thick AGN with a larger \nH\
than that given in Table 4, e.g., $\sim 10^{25}$ \psqcm.

Fe K emission is detected at $\sim 2\sigma$ or larger significance in
these objects except for PID 252 (Table 5, see also Comastri et al
2011). The spectrum of PID 252 does not show clear Fe K emission with
EW $\leq 0.5 $ keV ($2\sigma $ upper limit of a narrow line at 6.4
keV). This weak-line source may be a high-redshift analogue of Mrk
231, a Compton thick AGN with a weak Fe K line in the local Universe
(e.g., Braito et al 2004; Gallagher et al 2002; Iwasawa et al 2011 and
other references therein). The large EW observed in PID 114 (Table 5)
agrees with an Fe K line expected from a reflection-dominated spectrum
from cold medium, giving a support to
the possibility of a Compton thick source.

\begin{table}
\begin{center}
\caption{Fe K line equivalent widths of the V category objects.}
\begin{tabular}{rcc}
PID & EW$_1$ & EW$_2$ \\
 & keV & keV \\[5pt]
30 & $0.76\pm 0.24$ & $0.51\pm 0.19$ \\
64 & $0.57\pm 0.19$ & $0.27\pm 0.11$ \\
114 & $1.40\pm 0.61$ & $1.09\pm 0.53$ \\
144 & $1.13\pm 0.51$ & $0.47\pm 0.23$ \\
180 & $0.65\pm 0.16$ & $0.34\pm 0.12$ \\
245 & $1.10\pm 0.48$ & $0.44\pm 0.28$ \\ 
252 & $\leq 0.5$ & $\leq 0.2$ \\
\end{tabular}
\begin{list}{}{}
\item[] Note --- EW$_1$ and EW$_2$ are measured with respect to the
  rest-frame 5-10 keV continuum modelled by a simple power-law and an
  absorbed power-law of $\Gamma = 1.8$, respectively. For PID 252,
  $2\sigma $ upper limits are given. We remark that EW$_2$ are always
  smaller than EW$_1$ since part of the line is accounted for by the
  sharp continuum feature of an absorbed continuum carved by an Fe K
  edge.
\end{list}
\end{center}
\end{table}

\section{Discussion}

\subsection{X-ray selection of heavily obscured AGN}

Seven heavily obscured active galaxies with \nH $\thinspace\geq 0.6
\times 10^{24}$ \psqcm, including one previously known source (PID
144, Norman et al 2002; Comastri et al 2011), were selected by the
rest-frame X-ray colour selection, primarily utilizing the excess
emission in the 9-20 keV band relative to emission at lower
energies. Two of them (PID 114 and PID 252) are possibly Compton thick
AGN with a reflection-dominated spectrum. Given the limited bandpass
available from XMM-Newton, this selection can be applied only for high
redshift objects, but the a posteriori checks showed that this selection
is reliable for sources with spectra of reasonable quality, and can
pick up strongly absorbed sources with near Thomson-thick opacity.

This method is a pure X-ray selection, and these seven objects compose
a sample of heavily obscured, moderate-luminosity quasars with
$L_{10-20}\sim 10^{44}$ \ergps, selected by the hard X-ray emission
above 10 keV beyond the local Universe.

There are various reports in the literature on Compton thick AGN
candidates in the CDFS using whole or part of the Chandra 4 Ms data
(e.g., Norman et al 2002; Mainieri et al 2005; Tozzi et al 2006; Fiore
et al 2008; Gilli et al 2011; Feruglio et al 2011; Luo et al 2011;
Brightman \& Ueda 2012; Fiore et al 2012). Some of them lie in the
redshift range of our sample, although they are expectedly faint and
just a few of them entered in our sample of relatively bright
sources. Fiore et al (2012) investigated high-redshift sources at
$z>3$ in CDFS and selected several heavily obscured AGN. Their E537
(=PID 245), M5390 (=PID 144), M8273 (PID 180), M3320 (=PID107) and
M4302 (=PID 120) are in our sample. Our results on the spectra of
these sources agree except for PID 120 for which only moderate
absorption of \nH $\thinspace = 1.6^{+0.5}_{-0.6}\times 10^{22}$
\psqcm is found. The smaller $N_{\rm H}$ value for PID 245 obtained by
us (Table 4) is explained by the lower redshift adopted for this
source: the X-ray redshift $z_{\rm X}=2.68$ (Sect. 3.2, Table 3),
instead of the photometric redshift $z=4.29$ from the GOODS-ERS
(Grazian et al 2011) adopted by Fiore et al (2012), which a close
inspection of the X-ray/optical/infrared images suggests to be the
redshift for another galaxy near the X-ray source.

\subsection{Absorbed AGN fraction}

\begin{figure}
  \centerline{\includegraphics[width=0.42\textwidth,angle=0]{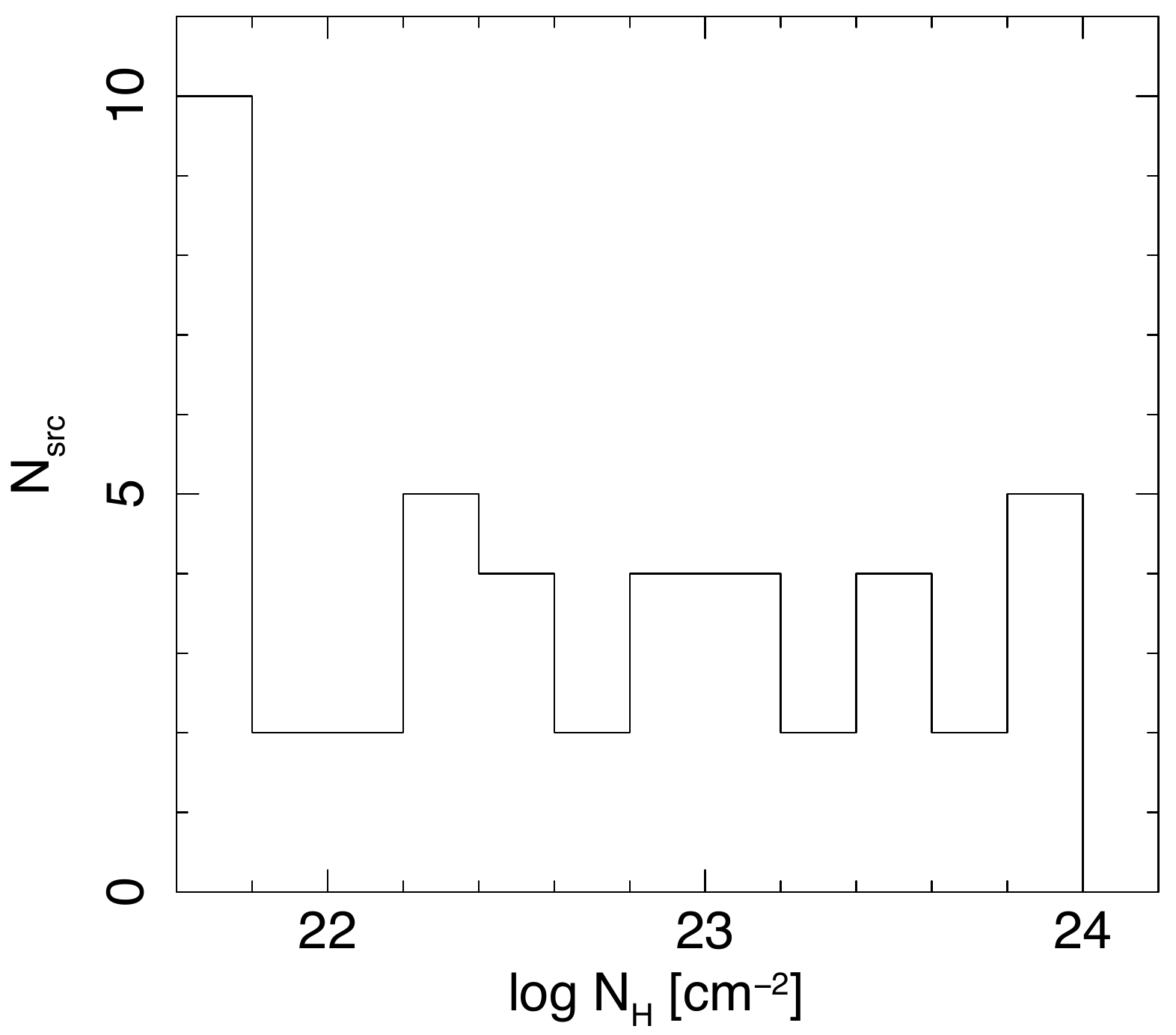}}
  \caption{The distribution of absorbing column density \nH, obtained
    by fitting an absorbed power-law to the EPIC spectra. The lowest
    bin represents the number of objects with no detection of
    absorption. The typical error bar of each bin is $\pm 1$.}

\end{figure}

In Fig.~3, when the model locus is used as a guide, 10 objects appear
to have {\it unobscured} X-ray sources, i.e., their X-ray absorption
is $N_{\rm H} < 10^{22}$ cm$^{-2}$. That is, $\sim 3/4$ of our sample
objects host signicantly obscured active nuclei. Fitting to the
individual spectra verifies the above assessment with 12 objects
having $N_{\rm H}$ values smaller than $10^{22}$ cm$^{-2}$, and gives
the $N_{\rm H}$ distribution shown in Fig.~7. The distribution of
log$N_{\rm H}$ (cm$^{-2}$) is nearly flat between 22-24, although the
two objects (PID 114 and 252) possibly move up to log $N_{\rm H}>24$
(cm$^{-2}$). For their typical 2-10 keV intrinsic luminosities
[(0.8-5)$\times 10^{44}$ erg s$^{-1}$], these active galaxies at
$z\sim 2.5$ can all be considered to emit at quasar luminosity, and
$74\pm 8$ per cent of them are absorbed X-ray sources. We have
estimated this absorbed AGN fraction using a Bayesian approach and the
binomial distribution (Wall \& Jenkins 2008) with a 68 per cent
confidence interval (S. Andreon, priv. comm.). It should be noted
that, since the fraction of Compton thick AGN is not constrained, this
value is considered to be the lower limit of the absorbed AGN
fraction.

We compared our findings with the predictions of the XRB synthesis
model by Gilli et al (2007). Our sample spans the 2-10 keV flux range
(2-54)$\times 10^{-15}$ erg cm$^{-2}$ s$^{-1}$. Since the sensitivity
of the XMM-CDFS observations strongly varies across the field, we
computed the model predictions at the 2-10 keV limiting flux,
$f^{lim}_{2-10}=4\times 10^{-15}$ erg cm$^{-2}$ s$^{-1}$, which
returns the same AGN surface density of our sample, i.e. 46 sources at
$z>1.7$ distributed over a $\sim0.27$ deg$^2$ area. The predicted
obscured fraction (defined as the number of AGN with log$N_{\rm H}>$22
over the total number of AGN in the sample) is $0.54\pm 0.06$,
smaller than the observed value of $0.74\pm 0.08$.

In the local Universe, the SWIFT/BAT and INTEGRAL surveys show that
absorbed sources (with \nH $\thinspace > 10^{22}$ \psqcm) consist
$\sim$55 per cent of hard X-ray selected AGN (e.g., Burlon et al 2011
and references therein). It is also found that this fraction depends
on X-ray luminosity, and at the luminosity matched to our sample, the
fraction is $21\pm 8$ per cent (Burlon et al 2011, see also Ebrero et
al 2008). The absorbed quasar fraction ($L^{\rm int}_{2-10}\geq
10^{44}$ erg s$^{-1}$) in our sample is higher than that of the local
Universe, suggesting a positive evolution with redshift, as found in
the previous work by La Franca et al (2005), Treister \& Urry (2006)
and Ebrero et al (2008). No evolution of the obscured AGN fraction was
assumed in Gilli et al (2007), yet the prediction comes close to the
observation at $z>1.7$ as discussed above. However, we note that the
luminosity dependence of the obscured AGN fraction assumed in Gilli et
al (2007) appears to be shallower than the observations (Hasinger
2008; Brusa et al 2010; Burlon et al 2011), and it overestimates the
obscured fraction of QSOs with $L_{\rm 2-10}>10^{44}$ \ergps\ in the
local Universe by a factor of $\sim 2.5$. This excess number assumed
for local obscured QSOs then compensates the lack of a redshift
evolution of their fraction in the model.

Contrary to the high-luminosity AGN, no strong evidence for a redshift
dependence of the obscured AGN fraction at luminosities $< 10^{44}$
\ergps\ has been found. Gilli et al. (2010), for instance, showed that
the increasing trend of the absorbed fraction {\it as observed} by
Hasinger (2008) for AGN with $L_{2-10} \leq 10^{44}$ \ergps, can be
accounted for by the K-correction effect, and is instead consistent
with a non-evolving {\it intrinsic} absorbed fraction. Here we suggest
that the obscured fraction increases with redshift {\it only} for
luminous QSOs. The different behaviours in obscured fraction between
low- and high-luminosity AGN may reflect their distinct accretion
mechanisms, as argued in literature (Hasinger 2008; Hopkins et al
2008; Hickox et al 2009): merger-driven accretion for luminous AGN
(e.g., Menci et al 2008) and secular accretion for less luminous AGN,
possibly mirroring their respective drivers of star formation (e.g.,
Elbaz et al 2011). This may not be the whole story but qualitatively
explains the different bahaviours between AGN of the low and high
luminosity ranges. If all QSOs originate from a major merger of
gas-rich galaxies (e.g., Sanders et al 1988), the increase of merger
rate at high redshift (with $\propto(1+z)^2$, e.g., Xu et al 2012)
naturally sees an increase in number of QSOs. A merger causes gas
channelling to the nuclear region (Barnes \& Hernquist 1991). This
concentration of gas and the chaotic geometry left by a merger would
lead to a high probability of the nuclear region to be seen obscured
(e.g., Hopkins et al 2006, but see Shawinski et al 2012) until the
radiation pressure of the buried QSO sweeps it away. In the context of
this evolutionary scenario alone, the obscured fraction of QSOs is
expected to be constant at all redshift, given the short duration of
the QSO lifetime ($\leq 10^8$ yr, Hopkins et al 2005). The evolution
we observed is probably driven by the increase in the gas fraction of
a galaxy towards high redshift (e.g., Carilli et al 2011), combined
with the efficient inflow induced by a merger. A higher gas fraction
of merger progenitor galaxies means more gas to be transported to the
nuclear region to form heavier obscuration. This would result in a
longer duration of the obscured phase, which can be translated to a
higher obscured fraction of the QSO population at high redshift. At
the same time, the elevated gas density by a merger increases the
efficiency of star formation leading to a starburst (e.g., Barnes \&
Hernquist 1991, Elbaz et al 2011). Kinetic energy injection from a
starburst may help to maintain the obscuration by inflating gaseous
wall around AGN (e.g., Fabian et al 1998).  Conversely, the lack of
mergers may explain the little evolution of the obscured fraction in
lower luminosity AGN. The gas fraction of galaxies hosting them also
increases towards high redshift in the same way as for high-luminosity
AGN. However, without a major merger, the gas reservoir is not
transported to the nuclear region rapidly. This means that the nuclear
obscuration condition remains little affected regardless the amount of
gas contained in a galaxy (hence redshifts). The gas content is
instead consumed to form stars over galaxy-wide as a secular process,
and the feeding to the black hole from a large-scale disk remains
relatively inefficient.

\bigskip

In summary, we present a result of a rest-frame 9-20 keV selection of
heavily obscured AGN at $z>1.7$, using the deep XMM-CDFS survey, and
also show that the fraction of absorbed AGN at high luminosity may be
higher at high redshift than in the local Universe. In the near
future, a further advance in this area of research will benefit from
even deeper observations of deep fields with Chandra and XMM-Newton,
while NuSTAR and Astro-H which will provide us with useful templates
and insights at lower redshifts. It is also useful to standardize
various X-ray spectral models of strongly absorbed systems with
improved physics incorporated for the community to share with.

\begin{acknowledgements}
  This research made use of the data obtained from XMM-Newton and the
  Chandra X-ray Observatory. KI thanks support from Spanish Ministerio
  de Ciencia e Innovaci\'on (MICINN) through the grant
  (AYA2010-21782-C03-01). WNB thanks the NASA ADP grant NNX10AC99G. We
  acknowledge financial contribution from the agreement ASI-INAF
  I/009/10/0.
\end{acknowledgements}


\begin{thebibliography}{}

\bibitem[Alexander et al.(2011)]{2011ApJ...738...44A} Alexander, D.~M., 
Bauer, F.~E., Brandt, W.~N., et al.\ 2011, \apj, 738, 44 

\bibitem[Alonso-Herrero et al.(2006)]{2006ApJ...640..167A} Alonso-Herrero, 
A., P{\'e}rez-Gonz{\'a}lez, P.~G., Alexander, D.~M., et al.\ 2006, \apj, 
640, 167 

\bibitem[Balestra et 
al.(2010)]{2010A&A...512A..12B} Balestra, I., Mainieri, V., Popesso, P., et al.\ 2010, \aap, 512, A12 

\bibitem[Barnes 
\& Hernquist(1991)]{1991ApJ...370L..65B} Barnes, J.~E., \& Hernquist, L.~E.\ 1991, \apjl, 370, L65 

\bibitem[Bauer et al.(2010)]{2010ApJ...710..212B} Bauer, F.~E., Yan, L., 
Sajina, A., \& Alexander, D.~M.\ 2010, \apj, 710, 212 

\bibitem[Biggs et al.(2011)]{2011MNRAS.413.2314B} Biggs, A.~D., Ivison, 
R.~J., Ibar, E., et al.\ 2011, \mnras, 413, 2314 

\bibitem[Braito et 
al.(2004)]{2004A&A...420...79B} Braito, V., Della Ceca, R., Piconcelli, E., et al.\ 2004, \aap, 420, 79 

\bibitem[Brandt et al.(1996)]{1996MNRAS.281L..41B} Brandt, W.~N., Iwasawa, 
K., \& Reynolds, C.~S.\ 1996, \mnras, 281, L41 

\bibitem[Brightman 
\& Ueda(2012)]{2012MNRAS.423..702B} Brightman, M., \& Ueda, Y.\ 2012, \mnras, 423, 702 

\bibitem[Brusa et al.(2010)]{2010ApJ...716..348B} Brusa, M., Civano, F., 
Comastri, A., et al.\ 2010, \apj, 716, 348 

\bibitem[Burlon et al.(2011)]{2011ApJ...728...58B} Burlon, D., Ajello, M., 
Greiner, J., et al.\ 2011, \apj, 728, 58 

\bibitem[Carilli et al.(2011)]{2011arXiv1105.1128C} Carilli, C.~L., Walter, 
F., Riechers, D., et al.\ 2011, arXiv:1105.1128 

\bibitem[Cooper et al.(2011)]{2011arXiv1112.0312C} Cooper, M.~C., Yan, R., 
Dickinson, M., et al.\ 2011, arXiv:1112.0312 

\bibitem[Cardamone et al.(2008)]{2008ApJ...680..130C} Cardamone, C.~N., 
Urry, C.~M., Damen, M., et al.\ 2008, \apj, 680, 130 

\bibitem[Comastri et 
al.(2011)]{2011A&A...526L...9C} Comastri, A., Ranalli, P., Iwasawa, K., et al.\ 2011, \aap, 526, L9 

\bibitem[Daddi et al.(2007)]{2007ApJ...670..173D} Daddi, E., Alexander, 
D.~M., Dickinson, M., et al.\ 2007, \apj, 670, 173 

\bibitem[Dahlen et al.(2010)]{2010ApJ...724..425D} Dahlen, T., Mobasher, 
B., Dickinson, M., et al.\ 2010, \apj, 724, 425 

\bibitem[Dickey 
\& Lockman(1990)]{1990ARA&A..28..215D} Dickey, J.~M., \& Lockman, F.~J.\ 1990, \araa, 28, 215 

\bibitem[Donley et al.(2012)]{2012ApJ...748..142D} Donley, J.~L., 
Koekemoer, A.~M., Brusa, M., et al.\ 2012, \apj, 748, 142 

\bibitem[Ebrero et 
al.(2009)]{2009A&A...493...55E} Ebrero, J., Carrera, F.~J., Page, M.~J., et al.\ 2009, \aap, 493, 55 

\bibitem[Elbaz et 
al.(2011)]{2011A&A...533A.119E} Elbaz, D., Dickinson, M., Hwang, H.~S., et al.\ 2011, \aap, 533, A119 

\bibitem[Fabian et al.(1998)]{1998MNRAS.297L..11F} Fabian, A.~C., Barcons, 
X., Almaini, O., \& Iwasawa, K.\ 1998, \mnras, 297, L11 

\bibitem[Feruglio et al.(2011)]{2011ApJ...729L...4F} Feruglio, C., Daddi, 
E., Fiore, F., et al.\ 2011, \apjl, 729, L4 

\bibitem[Fiore et 
al.(2012)]{2012A&A...537A..16F} Fiore, F., Puccetti, S., Grazian, A., et al.\ 2012, \aap, 537, A16 

\bibitem[Fiore et al.(2009)]{2009ApJ...693..447F} Fiore, F., Puccetti, S., 
Brusa, M., et al.\ 2009, \apj, 693, 447 

\bibitem[Fiore et al.(2008)]{2008ApJ...672...94F} Fiore, F., Grazian, A., 
Santini, P., et al.\ 2008, \apj, 672, 94 

\bibitem[Gallagher et al.(2002)]{2002ApJ...569..655G} Gallagher, S.~C., 
Brandt, W.~N., Chartas, G., Garmire, G.~P., 
\& Sambruna, R.~M.\ 2002, \apj, 569, 655 

\bibitem[George 
\& Fabian(1991)]{1991MNRAS.249..352G} George, I.~M., \& Fabian, A.~C.\ 1991, \mnras, 249, 352 

\bibitem[Giacconi et al.(2002)]{2002ApJS..139..369G} Giacconi, R., Zirm, 
A., Wang, J., et al.\ 2002, \apjs, 139, 369 

\bibitem[Gilli et 
al.(2007)]{2007A&A...463...79G} Gilli, R., Comastri, A., \& Hasinger, G.\ 2007, \aap, 463, 79 

\bibitem[Gilli et al.(2010)]{2010AIPC.1248..359G} Gilli, R., Comastri, A., 
Vignali, C., Ranalli, P., 
\& Iwasawa, K.\ 2010, X-ray Astronomy 2009; Present Status, Multi-Wavelength Approach and Future Perspectives, 1248, 359 

\bibitem[Gilli et al.(2011)]{2011ApJ...730L..28G} Gilli, R., Su, J., 
Norman, C., et al.\ 2011, \apjl, 730, L28 

\bibitem[Grazian et 
al.(2011)]{2011A&A...532A..33G} Grazian, A., Castellano, M., Koekemoer, A.~M., et al.\ 2011, \aap, 532, A33 

\bibitem[Guainazzi et 
al.(2000)]{2000A&A...356..463G} Guainazzi, M., Matt, G., Brandt, W.~N., et al.\ 2000, \aap, 356, 463 

\bibitem[Hasinger(2008)]{2008A&A...490..905H} Hasinger, G.\ 2008, \aap, 490, 905 
\bibitem[Hickox et al.(2009)]{2009ApJ...696..891H} Hickox, R.~C., Jones, 
C., Forman, W.~R., et al.\ 2009, \apj, 696, 891 

\bibitem[Hopkins et al.(2008)]{2008ApJS..175..356H} Hopkins, P.~F., 
Hernquist, L., Cox, T.~J., \& Kere{\v s}, D.\ 2008, \apjs, 175, 356 

\bibitem[Hopkins et al.(2006)]{2006ApJS..163....1H} Hopkins, P.~F., 
Hernquist, L., Cox, T.~J., et al.\ 2006, \apjs, 163, 1 

\bibitem[Hopkins et al.(2005)]{2005ApJ...625L..71H} Hopkins, P.~F., 
Hernquist, L., Martini, P., et al.\ 2005, \apjl, 625, L71 

\bibitem[Ikeda et al.(2009)]{2009ApJ...692..608I} Ikeda, S., Awaki, H., 
\& Terashima, Y.\ 2009, \apj, 692, 608 

\bibitem[Iwasawa et 
al.(2012)]{2012A&A...537A..86I} Iwasawa, K., Mainieri, V., Brusa, M., et al.\ 2012, \aap, 537, A86 

\bibitem[Iwasawa et 
al.(2011)]{2011A&A...529A.106I} Iwasawa, K., Sanders, D.~B., Teng, S.~H., et al.\ 2011, \aap, 529, A106 

\bibitem[Iwasawa et al.(1993)]{1993ApJ...409..155I} Iwasawa, K., Koyama, 
K., Awaki, H., et al.\ 1993, \apj, 409, 155 

\bibitem[La Franca et al.(2005)]{2005ApJ...635..864L} La Franca, F., Fiore, 
F., Comastri, A., et al.\ 2005, \apj, 635, 864 

\bibitem[Lehmer et al.(2005)]{2005ApJS..161...21L} Lehmer, B.~D., Brandt, 
W.~N., Alexander, D.~M., et al.\ 2005, \apjs, 161, 21 

\bibitem[Luo et al.(2011)]{2011ApJ...740...37L} Luo, B., Brandt, W.~N., 
Xue, Y.~Q., et al.\ 2011, \apj, 740, 37 

\bibitem[Luo et al.(2008)]{2008ApJS..179...19L} Luo, B., Bauer, F.~E., 
Brandt, W.~N., et al.\ 2008, \apjs, 179, 19 

\bibitem[Magdziarz 
\& Zdziarski(1995)]{1995MNRAS.273..837M} Magdziarz, P., \& Zdziarski, A.~A.\ 1995, \mnras, 273, 837 

\bibitem[Mainieri et al.(2005)]{2005MNRAS.356.1571M} Mainieri, V., 
Rigopoulou, D., Lehmann, I., et al.\ 2005, \mnras, 356, 1571 

\bibitem[Marconi et al.(2004)]{2004MNRAS.351..169M} Marconi, A., Risaliti, 
G., Gilli, R., et al.\ 2004, \mnras, 351, 169 

\bibitem[Mart{\'{\i}}nez-Sansigre et al.(2005)]{2005Natur.436..666M} 
Mart{\'{\i}}nez-Sansigre, A., Rawlings, S., Lacy, M., et al.\ 2005, \nat, 
436, 666 

\bibitem[Matt et 
al.(1999)]{1999A&A...341L..39M} Matt, G., Guainazzi, M., Maiolino, R., et al.\ 1999, \aap, 341, L39 

\bibitem[Matt et al.(1999)]{1999NewA....4..191M} Matt, G., Pompilio, F., 
\& La Franca, F.\ 1999, \na, 4, 191 

\bibitem[Matt et 
al.(1997)]{1997A&A...325L..13M} Matt, G., Guainazzi, M., Frontera, F., et al.\ 1997, \aap, 325, L13 

\bibitem[Menci et al.(2008)]{2008ApJ...686..219M} Menci, N., Fiore, F., 
Puccetti, S., \& Cavaliere, A.\ 2008, \apj, 686, 219 

\bibitem[Morrison 
\& McCammon(1983)]{1983ApJ...270..119M} Morrison, R., \& McCammon, D.\ 1983, \apj, 270, 119 

\bibitem[Murphy 
\& Yaqoob(2009)]{2009MNRAS.397.1549M} Murphy, K.~D., \& Yaqoob, T.\ 2009, \mnras, 397, 1549 

\bibitem[Nandra et al.(2007)]{2007MNRAS.382..194N} Nandra, K., O'Neill, 
P.~M., George, I.~M., \& Reeves, J.~N.\ 2007, \mnras, 382, 194 

\bibitem[Norman et al.(2002)]{2002ApJ...571..218N} Norman, C., Hasinger, 
G., Giacconi, R., et al.\ 2002, \apj, 571, 218 

\bibitem[Popesso et 
al.(2009)]{2009A&A...494..443P} Popesso, P., Dickinson, M., Nonino, M., et al.\ 2009, \aap, 494, 443 

\bibitem[Rafferty et al.(2011)]{2011ApJ...742....3R} Rafferty, D.~A., 
Brandt, W.~N., Alexander, D.~M., et al.\ 2011, \apj, 742, 3 

\bibitem[Sanders et al.(1988)]{1988ApJ...325...74S} Sanders, D.~B., Soifer, 
B.~T., Elias, J.~H., et al.\ 1988, \apj, 325, 74 

\bibitem[Santini et 
al.(2009)]{2009A&A...504..751S} Santini, P., Fontana, A., Grazian, A., et al.\ 2009, \aap, 504, 751 

\bibitem[Schawinski et al.(2012)]{2012MNRAS.425L..61S} Schawinski, K., 
Simmons, B.~D., Urry, C.~M., Treister, E., 
\& Glikman, E.\ 2012, \mnras, 425, L61 

\bibitem[Silverman et al.(2010)]{2010ApJS..191..124S} Silverman, J.~D., 
Mainieri, V., Salvato, M., et al.\ 2010, \apjs, 191, 124 

\bibitem[Szokoly et al.(2004)]{2004ApJS..155..271S} Szokoly, G.~P., 
Bergeron, J., Hasinger, G., et al.\ 2004, \apjs, 155, 271 

\bibitem[Taylor et al.(2009)]{2009ApJS..183..295T} Taylor, E.~N., Franx, 
M., van Dokkum, P.~G., et al.\ 2009, \apjs, 183, 295 

\bibitem[Tozzi et 
al.(2006)]{2006A&A...451..457T} Tozzi, P., Gilli, R., Mainieri, V., et al.\ 2006, \aap, 451, 457 

\bibitem[Treister et al.(2009a)]{2009ApJ...696..110T} Treister, E., Urry, 
C.~M., \& Virani, S.\ 2009, \apj, 696, 110 

\bibitem[Treister et al.(2009b)]{2009ApJ...693.1713T} Treister, E., Virani, 
S., Gawiser, E., et al.\ 2009, \apj, 693, 1713 

\bibitem[Treister 
\& Urry(2006)]{2006ApJ...652L..79T} Treister, E., \& Urry, C.~M.\ 2006, \apjl, 652, L79 

\bibitem[Vignali et al.(2010)]{2010MNRAS.404...48V} Vignali, C., Alexander, 
D.~M., Gilli, R., \& Pozzi, F.\ 2010, \mnras, 404, 48 

\bibitem[Vignati et 
al.(1999)]{1999A&A...349L..57V} Vignati, P., Molendi, S., Matt, G., et al.\ 1999, \aap, 349, L57 

\bibitem[]{} Wall, J.V., Jenkins, C.R., "Practical Statistics for Astronomers", p. 21, 2003, CUP

\bibitem[Wardlow et al.(2011)]{2011MNRAS.415.1479W} Wardlow, J.~L., Smail, 
I., Coppin, K.~E.~K., et al.\ 2011, \mnras, 415, 1479 

\bibitem[Wuyts et al.(2008)]{2008ApJ...682..985W} Wuyts, S., Labb{\'e}, I., 
Schreiber, N.~M.~F., et al.\ 2008, \apj, 682, 985 

\bibitem[Xu et al.(2012)]{2012ApJ...747...85X} Xu, C.~K., Zhao, Y., 
Scoville, N., et al.\ 2012, \apj, 747, 85 

\bibitem[Xue et al.(2011)]{2011ApJS..195...10X} Xue, Y.~Q., Luo, B., 
Brandt, W.~N., et al.\ 2011, \apjs, 195, 10 


\bibitem[Yaqoob(1997)]{1997ApJ...479..184Y} Yaqoob, T.\ 1997, \apj, 479, 
184 

\end{thebibliography}
\end{document}